\documentclass[aps, prl, reprint]{revtex4-1}
\usepackage{blindtext}

\usepackage[pdftex]{graphicx}%
\usepackage[version=4]{mhchem}%
\draft % marks overfull lines with a black rule on the right

\newcommand{\figref}[1]{\figurename~\ref{#1}}%

\begin{document}

\title{Direct evidence of jets emanating from droplets at the Rayleigh
  charge-induced instability point}

\author{Styliani Consta}
\email[]{sconstas@uwo.ca}
\affiliation{Department of Chemistry, The University of Western
  Ontario, London, Ontario, Canada N6A 5B7}
\affiliation{Yusuf Hamied Department of Chemistry, University of Cambridge,
Lensfield Road, Cambridge, CB2 1EW, United Kingdom}

\date{\today}

\begin{abstract}
  Highly charged liquid droplets are unstable above the critical charge
  squared-to-volume ratio given by the Rayleigh limit. 
  The instability leads to ion ejection from jets formed on the droplet's surface.
  Despite the many experiments that have been performed to capture the jet formation
  the precise fission
  mechanism has not yet been observed because of its brief
  transient nature.
  Here, we present the
  first atomistic simulations that reveal the mechanism of Rayleigh fission.
  We demonstrate that ion ejection
  takes place through a drop's deformation from a spherical into a
  distinct shape that contains a conical protrusion. We assert
  that the latter state is a free energy minimum along an order parameter
  that measures the degree of droplet asphericity.
  The charged droplet's long-time evolution proceeds
  by alternating between the two minima above and below the critical
  value that are reached through solvent evaporation and ion ejection,
  respectively.  For the first time, this mechanism allows one to explain
  the nature of the progeny droplets and the percentage of charge lost during
  fission.  We determine that the cone half-angle is close
  to the value predicted from the solution of the electrostatic equation
  for the dielectric liquid. It is found that the conical deformation is independent of
the effect of electrohydrodynamic forces reported in experiments.
 Contrary to the experimental observations
  of two diametrically opposite jets for droplets suspended in the
  electric field, we found that a single jet is
  formed at the Rayleigh limit. This indicates that super-charged droplet 
  states may have been detected in the experiments. 
\end{abstract}

% insert suggested PACS numbers in braces on next line

\maketitle %\maketitle must follow title, authors, abstract and \pacs

Charged liquid droplets are unstable above a critical charge
  squared-to-volume ratio given by the Rayleigh limit\cite{rayleigh1882}. This
  instability arises from competing electrostatic and surface tension forces.
  The instability leads to formation of jets ejecting ions from the droplet surface.
  The Rayleigh instability appears in numerous applications that include sprays used
  in native mass spectrometry, manufacturing, and
  inkjet printing\cite{jung2018three}. It also plays a role in
  phenomena of quantum mechanical nature\cite{lukyanchuk2015rayleigh,
    salomaa1981structure, Leiderer2015}.
  The fission of charged droplets has been studied for
  approximately one and a half centuries starting from the seminal article of 
  Lord Rayleigh\cite{rayleigh1882}.
Lord Rayleigh determined the conditions for the onset of instability in a charged
conducting droplet\cite{rayleigh1882}. 
 It is notable that in the same
article Lord Rayleigh intuited that the instability would subsequently
develop into a jet that emits ions.

Rayleigh's approach is based on a linear stability analysis of a
spherical droplet with respect to small shape perturbations expressed in
terms of spherical harmonics.  The energy of the droplet is written as
the sum of surface energy and electrostatic energy. This energy is
then expressed as a quadratic form in terms of the amplitudes of the
spherical harmonics used to describe the perturbations. For small
values of the charge the quadratic form is positive definite and,
hence, the droplet is stable. The condition of stability is concisely
described via the fissility parameter defined as
\begin{equation}
  \label{eq:fissility}
  X = \frac{ Q^2} {64 \pi ^2 \sigma \varepsilon _0 R_0^3},
\end{equation}
where $Q$, $R_0$, and $\sigma$ denote the total charge of the droplet,
the droplet radius, and the surface tension, respectively, and
$\varepsilon _0$ is the permittivity of vacuum. $X = 1$ corresponds to
the Rayleigh limit while for $X < 1$ the system is below the Rayleigh
limit and is stable w.r.t. small perturbations. 
The derivation of the Rayleigh limit following different approaches can be found in 
Refs.\cite{peters1980rayleigh, hendricks1963, consta2015disintegration}. 
The Rayleigh theory can not
provide an answer on the pathway of the droplet fission.  The mode
that corresponds to the order of the spherical harmonics ${l=2}$ is
the first to become unstable. It has been assumed that oblate-prolate
fluctuations, appearing as an hour-glass shape deformations, 
play a major role in the droplet fission \cite{nix1967normal}.

Following Lord Rayleigh's development, the ion ejection from charged droplets 
has been extensively tested experimentally \cite{hayati1986mechanism, Doyle1964Instability,
  Abbas1967Instability, schweizer1971stability, Richardson1989,
  delaMora1996, Smith2002, grimm2009evaporation, Taflin1989, Duft2002, Duft2003, achtzehn2005coulomb, 
gomez1994charge}. 
Many of the experiments have found the amount of charge released from a droplet and
some succeeded in capturing micro-graphs of the released train of droplets
from a jet\cite{Duft2002, Duft2003, gomez1994charge, Smith2002, grimm2009evaporation}.
However, the precise fission mechanism, which includes the birth and the retraction of the jet, 
has not been yet experimentally observed because of its brief
transient nature.  Additionally, in experimental setups the jet formation is influenced by aerodynamic
effects, impurities, and external electric fields\cite{gomez1994charge, Duft2002, Duft2003}, which
prevent a clear observation of the jets conjectured by Lord Rayleigh.
Quantitative explanations of the experimentally observed jet angles and of the jet stability 
are still missing from the literature.
The jets formed in conducting droplets and the multi-conical protrusions in the 
``star''-shaped 
super-charged dielectric droplets\cite{oh2017droplets} have not been related  
thus far.

In this Letter we present the first direct evidence via atomistic modeling of the 
jet formation at the Rayleigh limit.
Contrary to the experimental observations\cite{Duft2002, Duft2003, gomez1994charge}
of two diametrically opposite jets for droplets suspended in the
electric field, we find that a single jet is
formed at the Rayleigh limit. 

Extensive molecular dynamics (MD) simulations of atomistically modeled systems
of various sizes were performed.
It was estimated 
and empirically tested that 
a system of $3.2\times 10^4$ \ce{H2O} molecules  and 48 \ce{Na+} ions is sufficiently
large in order to produce realistic jets 
relevant to experimental observations.
The water molecules were represented by the TIP3P (transferable intermolecular 
potential with three
points)\cite{jorgensen1998temperature} water
molecules. The ions were modeled with the
CHARMM36m\cite{noskov08, beglov94} force field. The droplet was placed
in a spherical cavity with radius 20nm where the spherical boundary
conditions were imposed. The cavity size was chosen to accommodate the
extensive shape fluctuation of the droplet.  The simulations were
performed using the software NAMD version
2.12\cite{phillips05scalable}. Newton's equation of motion for each
atomic site was integrated using the velocity-Verlet algorithm with a
time step of 1fs. The temperature of
the system was set at 300K and was maintained with Langevin
thermostat with the damping coefficient set to 1/ps. The electrostatic
interactions were treated with the multilevel summation
method\cite{hardy2015multilevel}. The Rayleigh limit of the droplet
($X = 1$ in Eq.~\ref{eq:fissility}) was calculated with the surface
tension value\cite{vega2007} ${\sigma=52.3
\mathrm{mJ/m^2}}$. During the course of 1ns equilibration run no ions
were ejected from the system.  As the water molecules evaporate from
the surface of the drop the system crosses the instability boundary.
We observed the development of jets when the system size reached 28522
water molecules. We confirmed, by preparing systems comprising 28000
and 27500 water molecules, that the drops below this size also become
unstable and emit ions through jet formation.

As the droplet approaches the critical size (from below the Rayleigh threshold)
intermittent spines developed. The spines dimensions are comparable to
the diameter of the droplet, therefore, they cannot be considered
small perturbations of the spherical shape. At the critical size, that corresponds
to 28522 \ce{H2O} molecules and 48 \ce{Na+} ions, a stable cone
develops. 
The conical deformation is independent of
the effect of electrohydrodynamic forces reported in experiments\cite{melcher1969electrohydrodynamics,
hayati1986mechanism}. The observed geometry of the developed jets differ critically from the expected Taylor
cone geometry\cite{taylor1964}. 

\begin{figure}[htb!]
  \includegraphics[width=\linewidth]{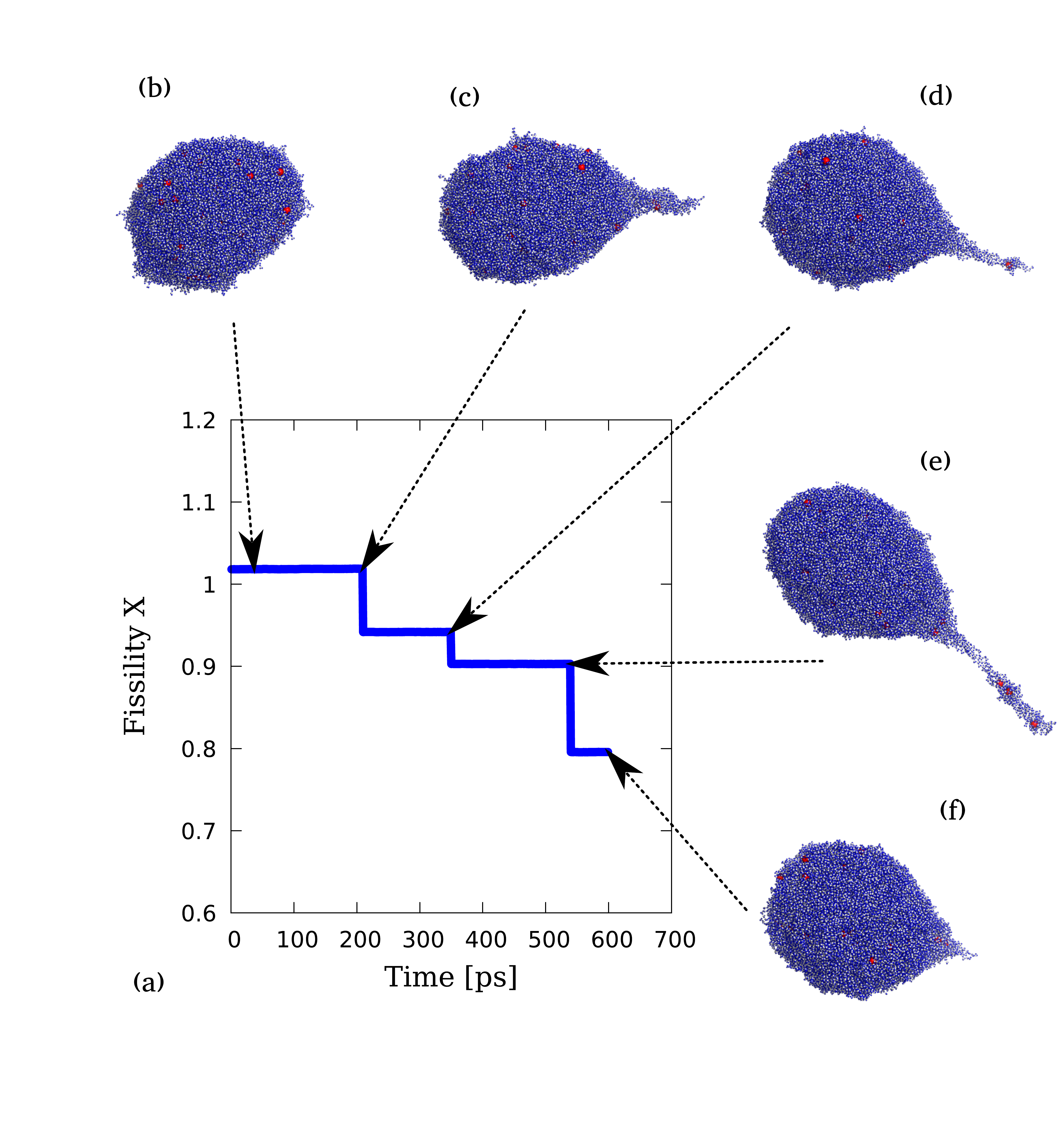}
  \caption{(a) Fissility parameter (Eq.~\ref{eq:fissility})
    in the course of the droplet
    fission. Only the connected droplets were used to compute the
    values of $X$.  The steps in the plot correspond to the
    fission events. (b)-(f) Typical snapshots of droplet states.
    The \ce{Na+} ions are shown by red spheres,
    while the \ce{H2O} molecules are blue colored.
    (b) A spherical
    conformation at the outset of the simulations.
    (c)-(e) Ion ejection events take place. In each ejection event
    different numbers of the ions are leaving the droplet. Two, one
    and three ions are being ejected at 190~ps, 340~ps and 530~ps time
    intervals, respectively. (f) Two ions are present at the base
    of the cone.  However, the cone collapsed before the ejection
    could take place.
  }
\label{fig:fragmentation-snapshots}
\end{figure}

\figref{fig:fragmentation-snapshots}(a) shows the time evolution of the fissility parameter $X$
in the course of the observed drop fragmentation. 
\figref{fig:fragmentation-snapshots}(b)-(f) are typical snapshots of
droplet states. The Supplementary Video shows the molecular dynamics
trajectory of the formation of the jet
and the ejection of the solvated ions. It is
emphasized that the spines are the result of the global fluctuation of
the charge density and can not be attributed to the local aggregation of
several charges within their bases.

\begin{figure}[htb!]
   \includegraphics[width=0.45\linewidth]{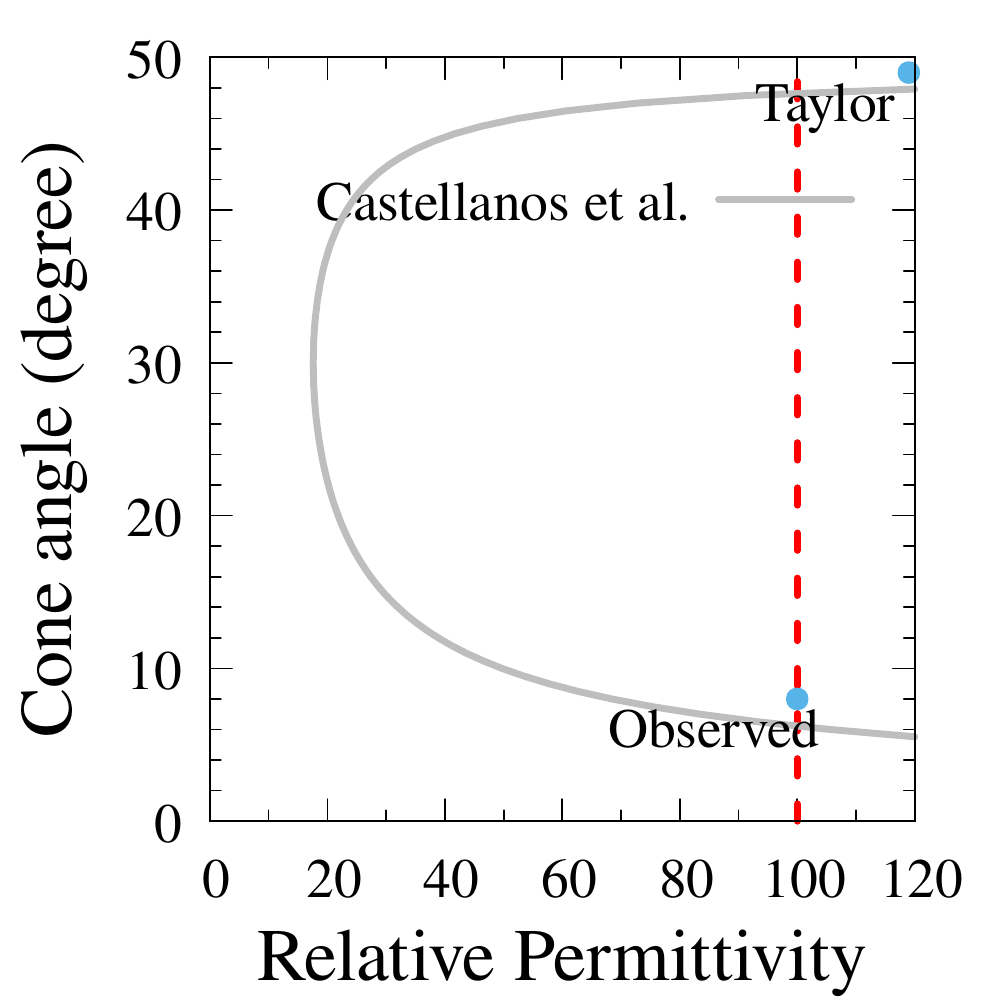}
  \includegraphics[width=0.45\linewidth]{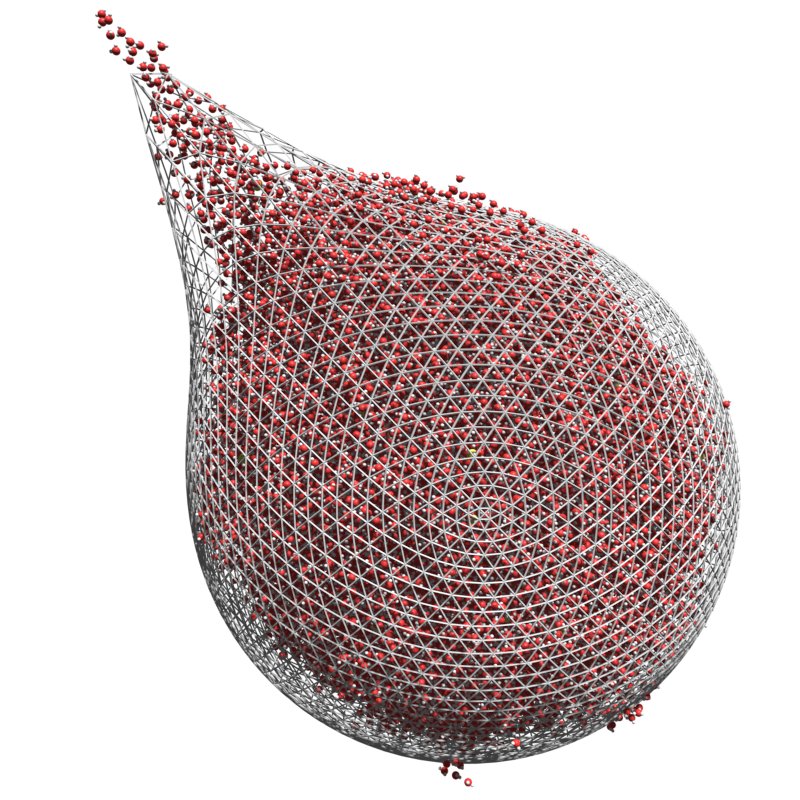}
  \caption{In the left panel the half-angle of the
    Taylor cone\cite{taylor1964, ramos1994conical}
    vs the dielectric permittivity of the solvent is plotted. The blue dots
    mark the Taylor-half angle and the observed half-angle angle
    for water. The Taylor half-angle
    ${49^o}$ corresponds to the case of the conductor. For the finite
    value of the dielectric constant two solutions are possible. The
    red line marks the value of the dielectric constant for TIP3P
    water model (dielectric constant\cite{zarzycki2020temperature} at $T=300$~K is 
    $\approx 100.9$). In the right panel a snapshot of a conformation comprising a
    conical jet is shown. The mesh is a numerical fit of the
    conformation to a spherical droplet with single conical
    protrusion. The fit is in agreement with the lower value of the
    solution (grey line in the left panel).
     }
  \label{fig:jet}
\end{figure}

Castellanos\cite{ramos1994conical} et al. provided an expression for the dielectric
constant that corresponds to the stable conical shape with half-angle
${\theta}$:
\begin{equation}
  \epsilon =
  \frac{P'_{\frac{1}{2}}(\cos\theta)P_{\frac{1}{2}}(-\cos\theta)}{P'_{\frac{1}{2}
    }(-\cos\theta)P_{\frac{1}{2}}(\cos\theta)}
  \label{eq:castelanos}
\end{equation}
where ${ P_{\frac{1}{2}} }$ is the Legendre function of degree
$\frac{1}{2}$. Eq. (\ref{eq:castelanos}) is an extension of the
Taylor solution for the conical drop deformation\cite{taylor1964, FernandezdelaMora2006}.
\figref{fig:jet} shows the values of the half angle as a function of
the relative permittivity. 
The upper and lower branches in \figref{fig:jet} show little sensitivity to the 
solvent's dielectric constant for values greater than $\approx 50$.   
Droplet configurations were fitted
using the approximation (\ref{eq:tear}) of the molecular 
surface that has the features of a tear-drop shape
\begin{equation}
  \begin{aligned}
    r(\phi,\theta) & = R + D e^{-\nu \psi} \\
    \cos \psi & = \sin \theta_0 \sin \theta \cos(\phi-\phi_0) + \cos
    \theta_0 \cos \theta
  \end{aligned}
  \label{eq:tear}
\end{equation}
where $\psi$ is the spherical angle between the direction of the
radius vector and the direction of the cone ${(\phi_0,
  \theta_0)}$. After some algebra one arrives at the corresponding
expression for the cone angle
\begin{equation}
  \tan \alpha/2 = \frac{R+D}{\nu D} .
  \label{eq:cone}
\end{equation}
The numerical fitting shows that the observed cone half-angle corresponds to the lower branch
of solutions with values for the half-angle of about
${\mathrm{10^\circ}}$ rather than ${\approx 49^\circ}$ suggested by
Taylor. Drops with similar acute conical deformations have been observed in 
experiments where they have been attributed to the
spatial charge from the emitted spray surrounding the droplet\cite{de1992effect}.

\begin{figure}[htb!]
\includegraphics[width=\linewidth]{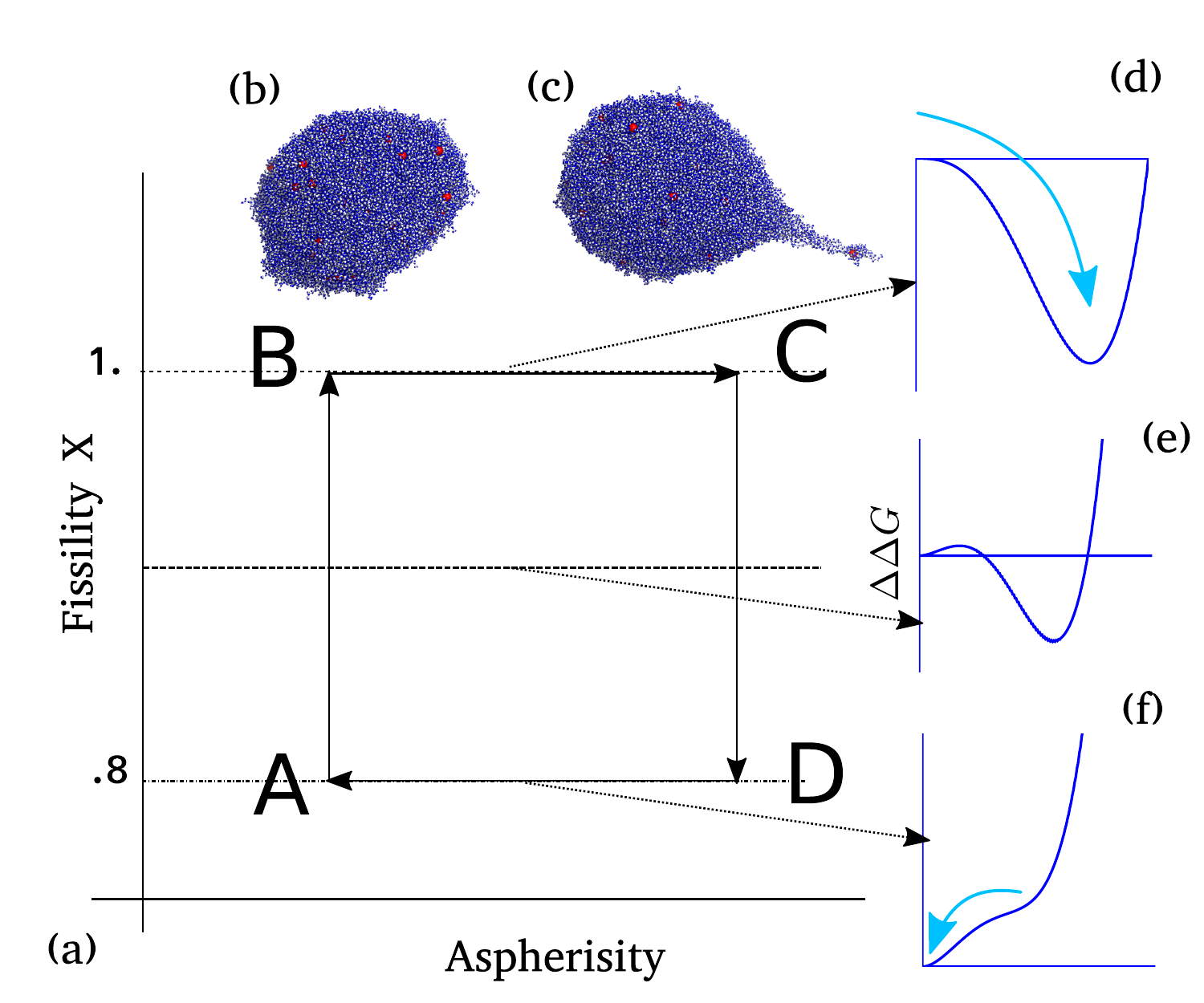}
  \caption{(a) Schematic illustration of the observed pattern of ion
    evaporation though a cyclic process. 
    (b)-(c) Typical spherical and aspherical droplet
    conformations. (d)-(f) Expected free energy profiles
    for the values of the fissility coefficients
    ${X=\{1.0, 0.9, 0.8\}}$.
    Free energies are plotted as a function of the deviation
    from the spherical shape. Details are discussed in the text.
    }
  \label{fig:fe}
\end{figure}

\figref{fig:fe}(a) shows the hypothetical cyclic evolution of a charged conducting
droplet characterized by the fissility ($X$) as a function of an order
parameter that measures the degree of deviation of the shape from the
sphere. We name this order parameter ``asphericity''.  
The droplet fission takes place by alternating between 
two minima in the free energy profile above and below the critical
value that are reached through solvent evaporation and ion ejection,
respectively.  State A is found at $X < 1$ and corresponds to the lowest
value of asphericity (\figref{fig:fe}~(b)).  As the drop evaporates
and $X$ increases, as it is inversely proportional to the size,
the stable spherical drop becomes a metastable conformation. In terms of the free energy
as a function of the asphericity, the meta-stability is demonstrated by a deep second
minimum. When the drop reaches state B where ${X=1}$ the drop becomes
unstable and evolves into a new state C that contains a conical
protrusion (\figref{fig:fe}~(c)). The ions enter the jet domain in a
diffusive motion and are ejected from the tip
(the details of the process can be viewed in the suppl. video). The electric field at
the tip diverges so that there is no activation energy barrier in this Born
ion ejection process. The system is locked in the new metastable state
while the charge ejection takes place. As more and more ions are emitted
$X$ decreases below a certain critical value at about ${X_{\mathrm{crit}}\approx{}.8}$ while the
cone is still present (State D).  At this time state D spontaneously
reverts to a new stable spherical state A. In smaller droplets low
activation barrier may allow for tunneling between states A and D for
relatively high values of the fissility parameter ${X > X_{\mathrm{crit}}}$. 
After this cycle the size of the droplet in the state A is smaller than that of in the original state A. 

We point out the presence of a hysteresis effect along the CD segment
in \figref{fig:fe}~(a). When the drop loses a number of charges and
the $X$ decreases below the critical value the fission process still
continues.  Here, we examined the largest droplet that can be modeled
atomistically and can demonstrate the formation of jets.  Small droplets
also show the formation of the tips, but the ion ejection is limited
to single ejection event comprising a single ion. In this case the
ejection event is indistinguishable from the Born model of ion
evaporation\cite{Iribarne1976, roux1990molecular}.

The ejection mechanism has a number of important consequences. We
conclude that the ions are ejected from the narrow conical
tip. Therefore, the size of the progeny droplets should not be overly
sensitive to the size of the parent droplets but may depend on the size of the ejected ion. 
The amount of the charge
lost should be a universal number for macroscopic drops.  In
microscopic drops there is a probability of tunneling between states D
and A making the amount of charge loss smaller. In native mass
spectrometry one of the hotly debated questions is how the macroions
are ejected from droplets of at least 100~nm in size. The macroions
trapped at the base of the cone will be ejected in the same way as
elementary ions carrying with them only the immediate environment. The
drop will preferentially eject the (macro)ions residing near the surface
of the drop.

\begin{figure}[htb!]
  \includegraphics[width=.4\linewidth]{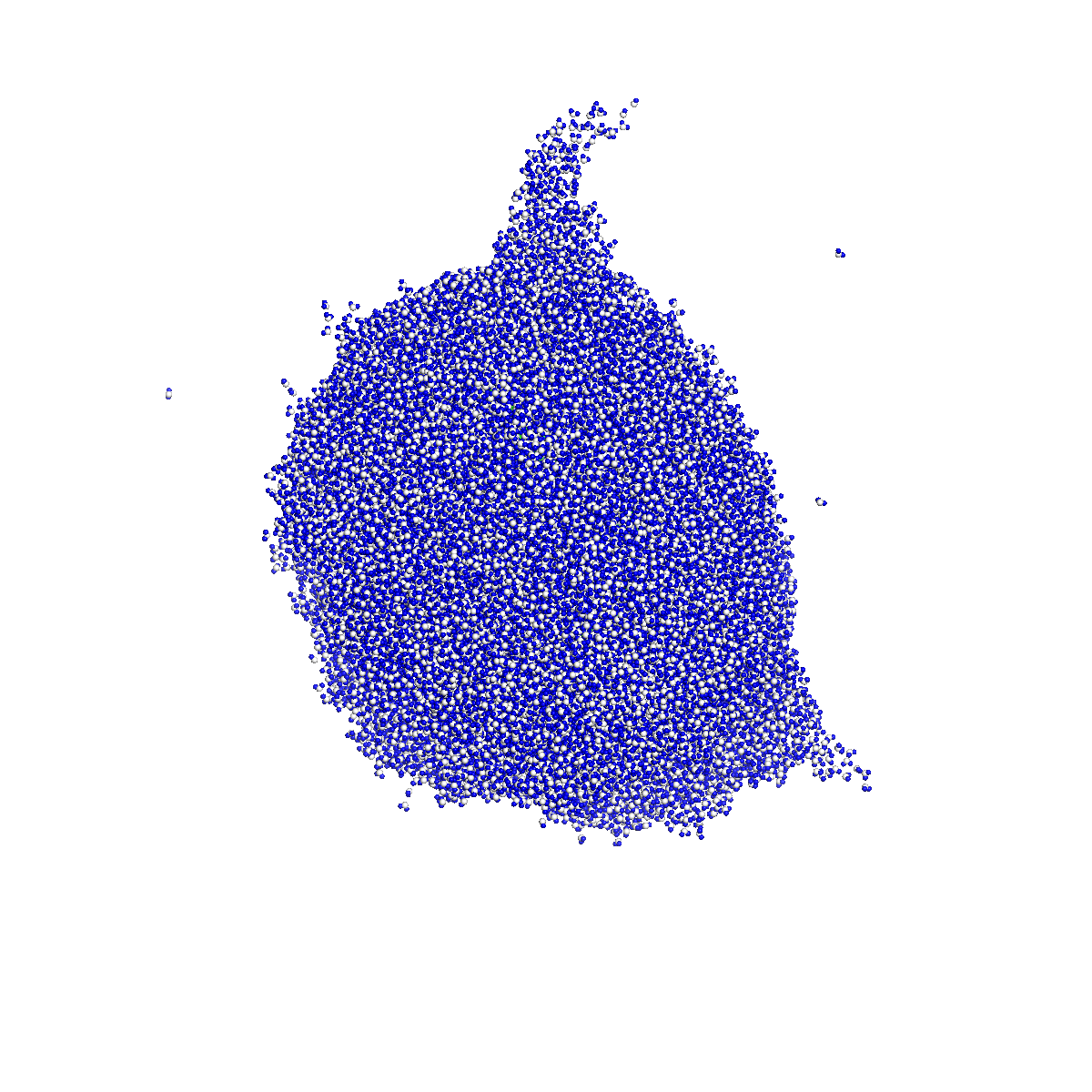}
  \includegraphics[width=.4\linewidth]{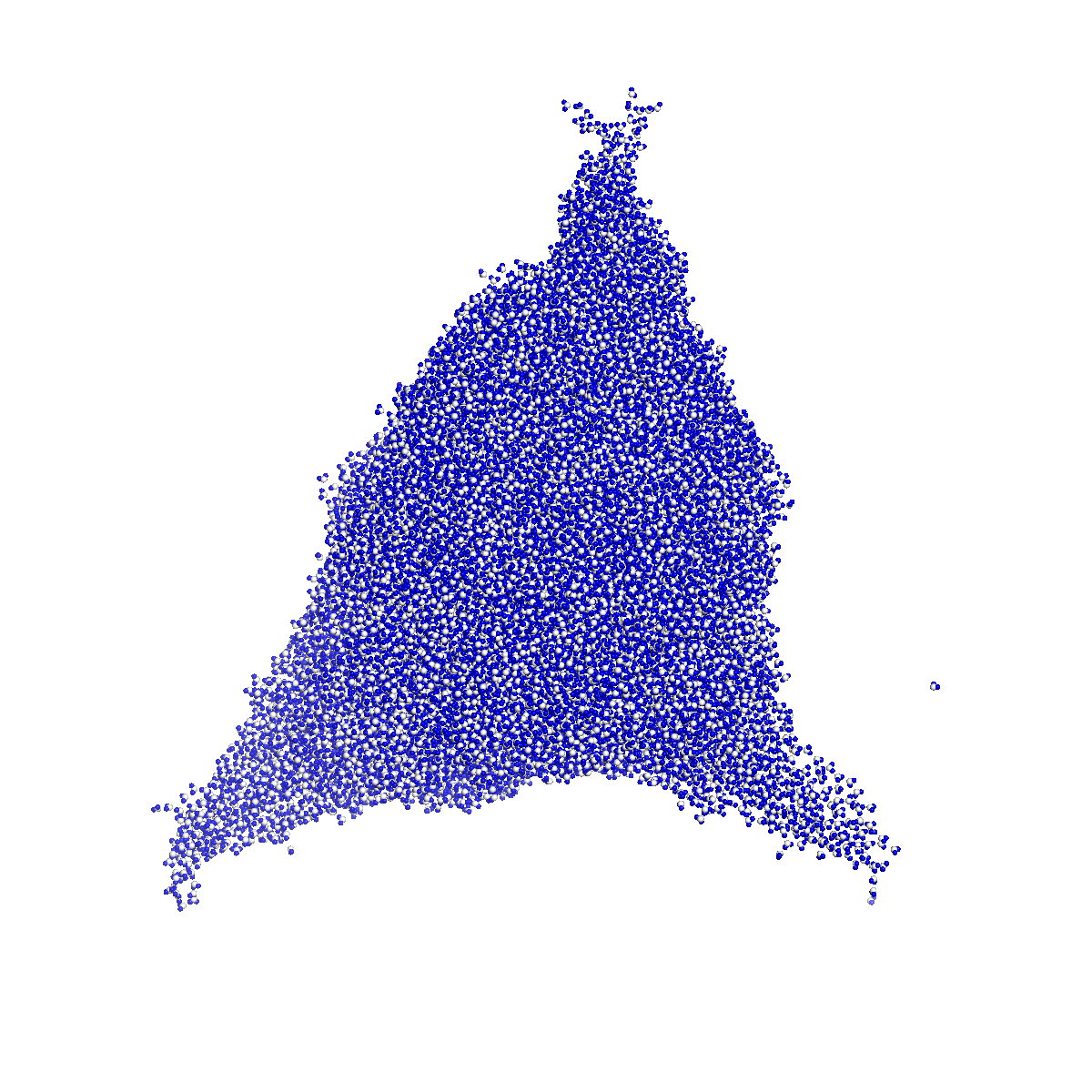}
  \caption{\label{fig:high-x}Stable (aqueous) droplet conformations that 
  contain a non-fissile charge in their center. These shapes
  are observed at high values of
    the fissility parameter (left) ${X=1.1}$ and (right) ${X=1.2}$.}
\end{figure}

An elegant explanation of the ejection mechanism that matches the
available computational evidence relies on the presence of an
asymmetric term in the free energy potential as illustrated in Ref.
\cite[pp. 334]{wales2004energy}. The variation of
the tentative free energy of the system along the asphericity is shown in panels
\figref{fig:fe}(d)-(f). The effect is similar to magnetization below
the Curie point. Here, the
deviation of the fissility parameter from the mean value plays 
the role of the applied magnetic field. When the
fissility parameter changes between the high and low values the system
stays trapped in one of the states as shown in
\figref{fig:fe}. Motivation for the presence of the odd terms in the
free energy potential comes from the statistical analysis of the
fluctuations of a droplet containing a non-fissile microion and was described in
detail in our prior work\cite{malevanets2018landau}. It is noted 
that charged dielectric droplets form ``star''-shapes above
the Rayleigh limit\cite{oh2017droplets} (Fig. \ref{fig:high-x}). To this end, we have
constructed a free energy functional for a weakly perturbed droplet
based on the general principles\cite{landau1980statistical}. We
postulated that the expansion of the free energy functional in powers
of the shape perturbation coefficients $a_{lm}$ should be invariant
with respect to the three dimensional rotations of the droplet and,
hence, should depend only on the Steinhardt
invariants\cite{steinhardt1983bond} $Q_{l}$ and
$W_{l}$.  $Q_l$ and $W_l$ parameters are proportional to the second
order and third order amplitudes of the expansion coefficients in
terms of spherical harmonics of order $l$.  The presence of the third
order invariant is rationalized on the basis of the observation of the
equilibrium star-shaped droplets. We notice that $Q^2_l$ terms are
invariant to the sign inversion transformation ${a_{lm} \to
-a_{lm}}$. However, an inspection of the shapes in \figref{fig:high-x}
shows that the stars have sharp cones pointing outwards and smooth
surface in the interior. This asymmetry indicates that the third order
invariant provides an important contribution to the free energy.  We
believe that the third order correction to the free energy functional
arises from a mean curvature correction to the macroscopic free
energy.  The presence of droplet protuberances on a smooth core
indicates that cones with positive mean curvature are favoured and
dimples in the droplet with negative mean curvature are penalized
explaining the observed droplet conformations.

The drop conformations comprising a single jet correspond to
quasi-static simulations when the evaporation is slower than the drop
dynamics. If the solvent evaporation is fast one may arrive at
super-charged drop with a value of the fissility parameter above
unity. In this case the drop fluctuations become significantly more
intensive and the multiple jet protrusions may develop. In
Fig. \ref{fig:high-x} snapshots of drop conformations with values of the
fissility parameter ${X=\{1.1,1.2\}}$ are presented. In this case the charges are
restrained to remain on a single non-fissile macroion to prevent
immediate drop disintegration. The simulations of the conducting and
charged dielectric droplets indicate that the experimentally
observed drops with multiple jets correspond to a case of
fissility parameter greater than one.  

In summary, we have provided direct evidence of the Rayleigh ejection mechanism
using atomistic modeling of charged drops. The modeling allowed us to
approach the Rayleigh limit very closely, which
cannot be readily achieved in experiments. The presented mechanism of jet formation
does not depend on the details of
the molecular models used. 
The formation
of a conical protrusion from which ions are released 
appears in the entire spectrum of droplet sizes ranging from the nano- to the
micro-size. In light of the proposed droplet morphology at the Rayleigh limit previous
experimental data need to be re-interpreted.
The atomistic insight into jet characteristics may lead to a unified theory 
that connects the instability in ferrofluids due to a magnetic field with the instability
manifested in ``star''-shaped droplets and jets emanating
from conducting droplets due to an electic field.

\begin{acknowledgments}
  S.C. thanks Prof.~D.~Frenkel, Department of Chemistry,
  University of Cambridge, UK and Dr. Anatoly Malevanets 
  for discussions on the stability of charged droplets.
  S.C.\ acknowledges the NSERC-Discovery
  grant for funding this research and a Marie Curie International
  Incoming Fellowship Grant Number 628552, held in the Department of
  Chemistry, University of Cambridge, UK.  ComputeCanada is
  acknowledged for providing the computing facilities.
\end{acknowledgments}

\end{document}